\documentclass[twocolumn,aps,pr,superscriptaddress]{revtex4-2}
\usepackage{amssymb} 
\usepackage{amsmath,bm} 
\usepackage{graphicx} 
\usepackage[normalem]{ulem} 
\usepackage{multirow} 
\usepackage[colorlinks,linkcolor=blue,urlcolor=blue,citecolor=blue]{hyperref} 
\usepackage{lipsum} 
\usepackage[usenames, dvipsnames]{xcolor} 
\usepackage{tensor} 
\usepackage{isotope} 
\usepackage{amsmath} 

\allowdisplaybreaks[4] 

\usepackage[colorlinks,linkcolor=blue,urlcolor=blue,citecolor=blue]{hyperref}
\makeatletter
\def\UrlAlphabet{%
\do\a\do\b\do\c\do\d\do\e\do\f\do\g\do\h\do\i\do\j%
\do\k\do\l\do\m\do\n\do\o\do\p\do\q\do\r\do\s\do\t%
\do\u\do\v\do\w\do\x\do\y\do\z\do\A\do\B\do\C\do\D%
\do\E\do\F\do\G\do\H\do\I\do\J\do\K\do\L\do\M\do\N%
\do\O\do\P\do\Q\do\R\do\S\do\T\do\U\do\V\do\W\do\X%
\do\Y\do\Z}
\def\UrlDigits{\do\1\do\2\do\3\do\4\do\5\do\6\do\7\do\8\do\9\do\0}
\g@addto@macro{\UrlBreaks}{\UrlOrds}
\g@addto@macro{\UrlBreaks}{\UrlAlphabet}
\g@addto@macro{\UrlBreaks}{\UrlDigits}
\makeatother

\newcommand{\dd}{\mathrm{d}}
\renewcommand{\sout}{\bgroup \color{red} \ULdepth=-.5ex \ULset}

\begin{document}

\title{Effects of chiral symmetry restoration on dilepton production in heavy ion collisions}

\author{Wen-Hao Zhou}
\email{zhou\_wenhao@fudan.edu.cn}
\affiliation{Key Laboratory of Nuclear Physics and Ion-beam Application (MOE), Institute of Modern Physics, Fudan University, Shanghai 200433}
\affiliation{Faculty of Science, Xi’an Aeronautical Institute, Xi’an 710077, China}
\affiliation{Shanghai Research Center for Theoretical Nuclear Physics, NSFC and Fudan University, Shanghai 200438, China}
\author{Che Ming Ko}
\email{ko@comp.tamu.edu}
\affiliation{Cyclotron Institute and Department of Physics and Astronomy, Texas A\&M University, College Station, Texas 77843, USA}
\author{Kai-Jia Sun}
\email{contact author:kjsun@fudan.edu.cn}
\affiliation{Key Laboratory of Nuclear Physics and Ion-beam Application (MOE), Institute of Modern Physics, Fudan University, Shanghai 200433}
\affiliation{Shanghai Research Center for Theoretical Nuclear Physics, NSFC and Fudan University, Shanghai 200438, China}

\begin{abstract}
Because of their weak interactions with the strongly interacting matter produced in relativistic heavy-ion collisions, dileptons provide an ideal probe of the early dynamics of these collisions. Here, we study dilepton production using a partonic transport model that is based on an extended Nambu-Jona-Lasinio (NJL) model. In this model, the in-medium quark masses decrease with increasing temperature as a result of the restoration of chiral symmetry. We find that the extracted temperature  from dileptons of intermediate masses agrees well with the temperature of the partonic matter, suggesting that dilepton production can be used as a thermometer for the produced partonic matter. Our results also indicate that the extracted in-medium quark masses decrease with increasing dilepton temperature, implying that dilepton production can further serve as a probe of chiral symmetry restoration in high energy heavy-ion collisions.
\end{abstract}
\maketitle

\section{Introduction}
\label{sec:introduction}
Because leptons interact weakly with the strongly interacting matter and are produced at all stages of relativistic heavy-ion collisions (HICs), they carry important information about the time evolution of the colliding systems~\cite{Linnyk:2015rco}. In HICs, dileptons (lepton pairs) can be produced from many sources~\cite{Kajantie:1986dh,Linnyk:2015rco,Drell:1970wh}, which include quark-antiquark annihilation in the initial partonic phase, meson-meson annihilation during subsequent hadronic phase, direct decays of neutral mesons ($\pi^0$, $\rho$, $\phi$, $J\!/\!\psi$ ...) and Dalitz decays of hadrons~\cite{Dalitz:1951aj} ($\eta$, $\omega$, $\Delta$ ...) after their kinetic freeze-out. Due to their diverse production sources, dileptons have been used to explore the in-medium properties of vector mesons~\cite{Li:1994cj,Bratkovskaya:2007jk,Metag:2007hq}, phase transition in quantum chromodynamics (QCD) ~\cite{Kajantie:1986dh,Islam:2014sea,Li:2016uvu,Savchuk:2022aev}, the strength of magnetic fields in HICs~\cite{Wei:2024lah,Das:2021fma,Wang:2022jxx} and the structures of atomic nuclei~\cite{Luo:2023syp,Lin:2022flv}. 

During the last two decades, numerous experimental measurements of dilepton production have been carried out by the STAR~\cite{STAR:2015tnn,STAR:2023wta,STAR:2024bpc,STAR:2015zal,STAR:2018ldd,Seck:2021mti} and PHENIX~\cite{PHENIX:2009gyd,PHENIX:2015vek} Collaborations at Relativistic Heavy Ion Collider (RHIC), the ALICE Collaboration~\cite{ALICE:2018ael} at the Large Hadron Collider (LHC), and the HADES Collaboration~\cite{HADES:2019auv} at GSI. These experiments have primarily focused on measuring the invariant mass spectrum of dileptons, which is typically divided into three regions: the low-mass region (LMR: $M_{ll}<M_\phi\sim$ 1 GeV), the intermediate-mass region (IMR: $M_\phi<M_{ll}<M_{J\!/\!\psi}\sim$ 3 GeV) and the high-mass region (HMR: $M_{ll}>M_{J\!/\!\psi}$). In HMR, dilepton production is well described by hadronic cocktail simulations~\cite{STAR:2015tnn}. However, in both LMR and IMR, the measured dilepton yields show significant enhancements over the cocktail contribution, and these have been mainly attributed to $\rho^0$-meson decay and quark-antiquark annihilation. Two types of theoretical models, namely effective many-body theory models~\cite{Rapp:2013nxa,Klusek-Gawenda:2018zfz,Galatyuk:2015pkq} and microscopic transport dynamic models (like PHSD~\cite{Linnyk:2011vx,Bratkovskaya:2012st} and UrQMD~\cite{Endres:2014zua,Endres:2016tkg}), have both been successful in explaining the data in the LMR by the broadened $\rho^0$ meson spectral function due to its rescattering in the expanding hadronic matter. In the IMR, the dilepton yield excess is mainly from emissions during the evolution of produced quark-gluon plasma (QGP)~\cite{STAR:2015tnn,Churchill:2023zkk,Churchill:2023vpt}.

Studies have suggested that dileptons can serve both as a thermometer and a chronometer of the strongly interacting matter produced in relativistic heavy-ion collisions~\cite{Rapp:2014hha,Churchill:2023zkk,Churchill:2023vpt}.	Recently, the STAR Collaboration has extracted the temperature of dileptons from their invariant mass spectra in Au+Au collision at $\sqrt{s_{NN}}=27$ and $54.4$ GeV~\cite{STAR:2024bpc}, and the results are $280\pm64\text{(stat.)}\pm10\text{(syst.)}\,\text{MeV}$ and $303\pm59\text{(stat.)}\pm28\text{(syst.)}\,\text{MeV}$, respectively, with large uncertainties due to limited statistics and the large background from hadronic cocktail contributions.  These temperatures  are much higher than the chemical freeze-out temperatures extracted from the particle yields measured in experiments~\cite{Andronic:2017pug,STAR:2017sal} and the chiral crossover transition temperature between the QGP and the hadronic matter from lattice QCD calculations~\cite{Aoki:2006we,Bazavov:2011nk}. 

In the present study, we investigate the production of dileptons in relativistic heavy ion collisions by using a partonic transport model with the interactions among partons described by an extended Nambu-Jona-Lasinio (NJL) model~\cite{Nambu:1961tp,Nambu:1961fr}. In this model, the in-medium quark masses gradually increases as a result of the spontaneous breaking of chiral symmetry when the produced partonic matter expands and cools. Extracting the temperature and quark effective masses from the IMR of the dilepton spectrum, we find that the dilepton temperature aligns well with the temperature of the partonic matter and the quark masses decrease with increasing dilepton temperature. These results suggest that dileptons can act as a thermometer for the produced partonic matter and also as a probe of chiral symmetry restoration in high energy heavy-ion collisions.

\section{Theoretical framework}
\label{sec:theory}

\subsection{Three-flavor Nambu–Jona-Lasinio model}
\label{sec:njl}
To describe the interactions in a partonic matter, we adopt the three-flavor NJL model with the following Lagrangian density~\cite{Nambu:1961tp,Nambu:1961fr,Buballa:2003qv},
\begin{align}
	\mathcal{L}_{\mathrm{NJL}} 
	=\;& \bar{\psi}(i\gamma^\mu\partial_\mu-\hat{m})\psi\notag\\
	&+\frac{G_{S}}{2}\sum_{a=0}^{8}[(\bar{\psi}\lambda_a\psi)^2+(\bar{\psi}i\gamma_5\lambda_a\psi)^2]
	\notag\\
	&-\frac{G_{V}}{2}\sum_{a=0}^{8}[(\bar{\psi}\gamma_\mu\lambda_a\psi)^2+(\bar{\psi}\gamma_5\gamma_\mu\lambda_a\psi)^2]
	\notag\\
	&-K\{\mathrm{det}_f[\bar{\psi}(1+\gamma_5)\psi]+\mathrm{det}_f[\bar{\psi}(1-\gamma_5)\psi]\}.
\end{align}
In the above, $\psi=\mathrm{diag}\left(u,d,s\right)$ is the three-flavor quark field; $\hat{m} = \mathrm{diag}\left(m_u,m_d,m_s\right)$ is the current quark mass matrix in the flavor space; $\lambda_a$ are the SU(3) flavor Gell-Mann matrices and $\lambda_0=\sqrt{2/3}I$; and the $G_S$, $G_V$ and $K$ are, respectively, the strengths of the scalar coupling, the vector coupling, and the axial $U(1)_A$ symmetry breaking Kobayashi-Maskawa-t’Hooft (KMT) interaction (six-fermion interaction). We note that a nonzero vector coupling has been shown to be needed in the NJL-based partonic transport model for describing the measured proton directed flow~\cite{Guo:2018kdz} and elliptic flow splitting of many identified hadrons at the beam energy scan experiments at RHIC~\cite{Ko:2013nbt,Xu:2013sta,Liu:2019ags,Zhou:2021ruf}. In the present study, we employ the parameters $m_u=m_d=5.5$ MeV, $m_s=140.7$ MeV, $G_S\Lambda^2=3.67$, $K\Lambda^5=12.36$ and a ultraviolet momentum cut-off $\Lambda=602.3\,\mathrm{MeV\!}/c$ for the momentum integrals in the NJL model as in Refs.~\cite{Buballa:2003qv,Hatsuda:1994pi}.

In the mean-field approximation, the NJL-Lagrangian for a quark matter at finite baryon chemical potential and temperature becomes
\begin{equation}\label{equ:MF}
	\mathcal{L}_\mathrm{MF} = \bar{\psi}(i\gamma^\mu\partial_\mu+\gamma^0\tilde{\mu}-\hat{M})\psi - \mathcal{V}.
\end{equation}
In the above,
\begin{equation}
	\mathcal{V}= G_S\left(\sigma_u^2+\sigma_d^2+\sigma_s^2\right) - 4K\sigma_u\sigma_d\sigma_s
	-\frac{1}{2}g_V\rho^2,\label{equ:potential}
\end{equation}
is the potential energy density with $g_V=\frac23 G_V$, $\hat{M}=\mathrm{diag}\left(M_u,M_d,M_s\right)$ is the effective mass matrix of quarks with
\begin{align}
	M_i =\;& m_i - 2G_{S}\sigma_i + 2K\sigma_j\sigma_k, \quad i\ne j\ne k,\label{equ:mass}
\end{align}
and $\tilde{\mu}=\mathrm{diag}\left(\tilde{\mu}_u,\tilde{\mu}_d,\tilde{\mu}_s\right)$ is the effective quark chemical potential matrix with
\begin{equation}
	\tilde{\mu}_i = \mu_i - g_{V}\rho_0,\label{equ:chemical}
\end{equation}
where the quark chemical potentials $\mu_i$ are taken to be $\mu_u=\mu_d=\mu$ and $\mu_s=0$ in this study.

In Eqs.~\eqref{equ:potential} and \eqref{equ:mass}, the quark condensates $\sigma_i$ with $i=u,d,s$, which are the order parameters of the chiral phase transition, are given by the integral,
\begin{equation}
	\sigma_i =\langle\bar\psi_i\psi_i\rangle= -2N_c\int_0^{\Lambda}\frac{\mathrm{d}^3\vec k}{\left(2\pi\right)^3}
	\frac{M_i}{E_i}\left(1-f_i-\bar f_i\right), \label{eq:gap}
\end{equation}
where $2N_c$ represents the spin and color degeneracy of the quarks and $E_i=\sqrt{M^2_i+\vec{k}}$ is the quark energy.  The $f_i$ and $\bar{f}_i$ in the above equation are the quark and antiquark phase-space distributions, respectively, and they are given by the Fermi-Dirac distributions,
\begin{equation}
	f_i = \frac{1}{1+e^{\beta(E_i-\tilde{\mu}_i)}},\quad\bar{f}_i = \frac{1}{1+e^{\beta(E_i+\tilde{\mu}_i)}},\label{equ:fdd}
\end{equation}
where $\beta=1/T$ with $T$ being the temperature of the partonic matter.

The net-quark number density $\rho$ in Eqs.~\eqref{equ:potential} and \eqref{equ:chemical} denotes the sum of net-$u$, net-$d$, and net-$s$ quark densities, i.e., $\rho=\rho_u+\rho_d+\rho_s$ with each being the zeroth component of its net-quark current density, which is given by
\begin{equation}
	\rho_i^\mu =\langle\bar\psi_i\gamma^\mu\psi_i\rangle= 
	2N_c\int_0^{\Lambda}\frac{\mathrm{d}^3\vec k}{\left(2\pi\right)^3}
	\frac{k^{\mu}}{E_i}\left(f_i-\bar f_i\right), \label{equ:density}
\end{equation}
where $N_c=3$ denotes the color of a quark.

From Eq.~\eqref{equ:MF}, the energy density of a quark matter can be obtained as
\begin{gather}
	\varepsilon_{\mathrm{NJL}}
	=-2 N_c\sum_{i=u, d, s}\int_0^{\Lambda} \frac{\mathrm{d}^3\vec k}{(2\pi)^3} E_i\left(1-f_i-\bar f_i\right)
	\notag\\
	+\;G_S\left(\sigma_u^2+\sigma_d^2+\sigma_s^2\right) 
	-4 K \sigma_u\sigma_d\sigma_s + \frac{1}{2}g_V\rho^2-\varepsilon_0,\label{equ:energy}
\end{gather}
where $\varepsilon_0$ is the vacuum energy density calculated by setting $f=\bar f=0$ in the above equation.

\subsection{Equations of motion}
\label{sec:eos}
In the semi-classical approximation, the single-particle Hamiltonian for a quark in the NJL model is given by~\cite{Ko:2012lhi},
\begin{equation}
	H = \sqrt{M^2+\vec{k}^2} \pm A^0.
\end{equation}
where $M$ denotes the in-medium quark mass given in Eq.(\ref{equ:mass}), $\vec{k}=\vec{p}\mp\vec{A}$ is the kinetic momentum of the quark with  $\vec{p}$ being its canonical momentum and ${\vec A}$ being the spatial components of the effective vector field acting on this quark, which together with the time component $A^0$ of the vector field are given by $A^\mu=g_V\rho^\mu$. The upper and lower signs in the above equation are for quarks and antiquarks, respectively. The equations of motion (EOM) of the quark can then be obtained from the Hamilton's equation of motion as
\begin{gather}
	\frac{\mathrm{d}\vec{r}}{\mathrm{d}t}=\frac{\vec{k}}{\sqrt{M^2+{\vec k}^2}}=\vec{v},\label{equ:drdt}\\
	\frac{\mathrm{d}\vec{k}}{\mathrm{d}t}=-\frac{M}{E}\vec{\nabla}M\pm\left(\vec{E}+\vec{v}\times\vec{B}\right),\label{equ:dkdt}
\end{gather}
where the effective electromagnetic fields are
\begin{gather}
	\vec{E}=-\vec{\nabla}A^0-\frac{\partial\vec{A}}{\partial t}, \\
	\vec{B}=\vec{\nabla}\times\vec{A}.
\end{gather}

\subsection{Dilepton production rate}
\label{sec:rate}
Dilepton production in a QGP mainly comes from the annihilation of quark and antiquark of same flavor, i.e., $q\bar q\rightarrow e^+e^-$, with a cross section given by~\cite{Peskin:1995ev},
\begin{equation}
	\sigma_{q\bar q\rightarrow e^+e^-} = \frac{4\pi\alpha^2e_{q}^2}{9s}, \label{equ:Si1}
\end{equation}
where $s$ is the square of the center-of-mass energy of the quark-antiquark pair or the invariant mass of produced dilepton, $e_{q}$ is the charge of the quark, and $\alpha=1/137$ is the fine-structure constant.

The dilepton production rate $\dd N_{q\bar q}^{ee}/\dd t$ from quark-antiquark annihilation in a quark-gluon plasma of volume $V$ is then given by~\cite{Gondolo:1990dk,Cannoni:2013bza}
\begin{align}
\frac{\dd N_{q\bar q}^{ee}}{\dd t} =\;&
  V\!\!\sum_{i=u,d,s}\tilde\rho_i\tilde\rho_{\,\bar i}\int_{0}^{\infty}\dd^3\vec k_1\dd^3\vec k_2 v_\mathrm{rel}\notag\\
  &\times\tilde{f}_i(\vec k_1)\tilde{f}_{\,\bar i}(\vec k_2)\sigma_{i\bar i\to e^+e^-}(\sqrt{s}), \label{equ:Ncoll0}
\end{align}
In the above, $\tilde\rho_i$ and $\tilde\rho_{\,\bar i}$ are the densities of quarks of flavor $i$ and antiquarks of flavor $\bar i$, respectively, and they are given by $\tilde\rho_{i,\bar i}=2N_c\int\frac{d^3\vec k}{(2\pi)^3}f_{i,\bar i}$; $\vec k_1$ and $\vec k_2$ are the momentum of the quark and antiquark, respectively; $\tilde f_i$ and $\tilde f_{\,\bar i}$ are the normalized quark and antiquark momentum distributions; and $v_\mathrm{rel}$ is the Lorentz invariant Møller velocity between the quark and antiquark, which is given by
\begin{equation}
	v_\mathrm{rel} = \sqrt{|\vec{v}_1-\vec{v}_2|^2-|\vec{v}_1\times\vec{v}_2|^2}
	= \frac{\sqrt{s(s-4m^2)}}{2E_1E_2},
\end{equation}
with $\vec{v}_{1,2}=\vec{k}_{1,2}/E_{1,2}$ and $m$ being the mass of the quark and its antiquark. 

Taking quarks and antiquarks to have a Maxwell-Boltzmann distribution in energy, which is valid for $E \pm\tilde{\mu}_i\gg T$, their normalized distribution is then given by
\begin{equation}
	\tilde{f}(\vec k) = \frac{1}{4\pi m^2TK_2(m/T)}
	\exp{\left(-\frac{E}{T}\right)}, \label{equ:fpp}
\end{equation}
with $K_1$ and $K_2$ being the modified Bessel functions of first and second kinds, respectively. With the above expression, Eq.~\eqref{equ:Ncoll0} can be expressed as
\begin{align}
	\frac{\dd N_{q\bar q}^{ee}}{\dd t}
	=\;& \frac{\rho^2V}{8m^4TK_2^2(m/T)} \notag\\& \times
	\int_{2m}^{\infty}\dd M\, M^2(M^2-4m^2)K_1\left(\frac{M}{T}\right)\sigma(M) \notag\\ \label{equ:Ncoll1}
	=\;&\frac{\pi\alpha^2e_q^2\rho^2V}{18m^4TK_2^2(m/T)}
	\!\int_{2m}^{\infty}\!\dd M(M^2-4m^2)K_1\!\left(\frac{M}{T}\right).
\end{align}
The rate for the production of dileptons of invariant mass $M$ is thus given by
\begin{equation}
	\frac{\dd N_{q\bar q}^{ee}}{\dd M \dd t} \propto 
	\frac{M^2-4m^2}{m^4TK_2^2(m/T)}
	K_1\left(\frac{M}{T}\right). \label{equ:fit}
\end{equation}
For dilepton of invariant mass $M$ and QGP of temperature $T$ satisfying $m\ll T\ll M$, the above formula can be approximated as
\begin{equation}
	\frac{\dd^2 N_{q\bar q}^{ee}}{\dd M\dd t} \propto 
	\frac{1}{T^{9/2}}M^{3/2}e^{-\frac{M}{T}}. \label{equ:qqee2}
\end{equation}
As suggested in Refs.~\cite{Rapp:2014hha,HADES:2019auv}, the temperature of QGP can be determined by fitting the dilepton invariant mass spectrum with a fitting function similar to Eq.~\eqref{equ:qqee2}. In the present study, we will employ the more accurate Eq.~\eqref{equ:fit} to extract the temperature of the QGP as well as the in-medium quark mass from the dilepton invariant mass spectrum.

\subsection{Scattering criterion}
\label{sec:collision}
For a more accurate description of the dynamic evolution of a partonic matter, we use the stochastic collision criterion~\cite{Xu:2004mz} to treat the elastic scattering between quarks and the test particle method~\cite{Wong:1982zzb} to solve for the phase-space distributions $f_i$ and $\bar f_i$ of quarks and antiquarks in the transport model. In this method with $N_{\rm test}$ test (anti)quarks for a physical (anti)quark, the scattering probability for a pair of (anti)quarks in a certain volume $\Delta V=\Delta x\Delta y\Delta z$ during a time interval $\Delta t$ is given by~\cite{Xu:2004mz}
\begin{equation}
	P_c=v_\mathrm{rel}\frac{\sigma}{N_\mathrm{test}}\frac{\Delta t}{\Delta V},
\end{equation}
where $\sigma$ is the scattering cross section between (anti)quarks, which we take to be isotropic and have the constant value of 10 mb. Particles are only allowed to scatter with one another within the same spatial cell. Since the process $q\bar q\rightarrow e^+e^-$ is a rare process, dilepton production is treated perturbatively in our study with the number of dileptons produced from an elastic scattering between a quark and antiquark pair given by the ratio $\sigma_{q\bar q\rightarrow e^+e^-}/\sigma$~\cite{Li:1994cj}.

To ensure thermally equilibrated quarks and antiquarks to have the Fermi-Dirac distributions given in Eq.~\eqref{equ:fdd}, the Pauli blocking effect is included in a (anti)quark-(anti)quark scattering with the blocking probability given by
\begin{equation}
	P_{b} = 1 - (1-f_1(\vec{r}_1,\vec{k}_1))(1-f_2(\vec{r}_2,\vec{k}_2)),
\end{equation}
where $\vec{r}_i$, $\vec{k}_i$ and $f_i$ are the coordinates, momenta and local (anti)quark distribution functions, respectively, for the two scattering (anti)quarks. For thermodynamic quantities, such as the temperature and chemical potentials in the local quark distribution functions $f_i$, they are obtained from fitting the local net-quark and energy densities. In this study, we set $\Delta x=\Delta y=\Delta z=1\;\mathrm{fm}$ and $\Delta t=0.1\;\mathrm{fm\!/}\!c$, and this parameter set ensures that the collision probability $P_c$ is always less than one.

\section{Results and discussions}
\label{sec:result}
In this section, we compare the temperature $T_{\rm Dilepton}$ extracted from the dilepton invariant mass spectrum to the temperature $T_{\rm Medium}$ of the partonic matter. For $T_{\rm Dilepton}$, it is determined from fitting different regions of the dilepton invariant mass spectrum together with the in-medium quark masses by using Eq.~\eqref{equ:fit}. The $T_\mathrm{Medium}$ is , on the other hand, obtained directly from the weighted mean temperature of the system over many small cells,
\begin{equation}
	T_\mathrm{Medium} = \frac{\sum_n w_n T_n}{\sum_n w_n}, \label{equ:ther}
\end{equation}
where $n$ sums over the cells of volume $\Delta V$ as used in Sec.~\ref{sec:collision}, $w_n$ denotes the product $\rho_n\bar\rho_n$ of quark and antiquark densities in cell $n$ to reflect the production strength of dileptons, and $T_n$ is the local temperature of this cell, which can be determined from the equation of state of the partonic matter by assuming local thermal equilibrium.
\begin{figure}[!t]
	\centering
	\includegraphics[width=0.48\textwidth]{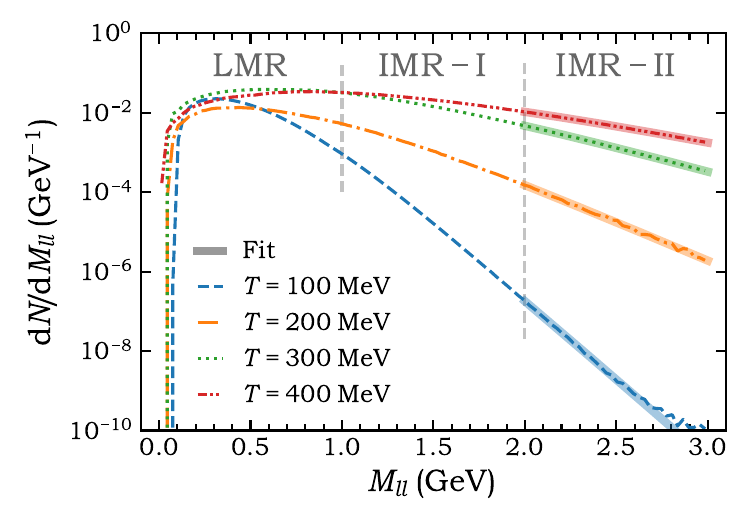}
	\caption{Invariant mass spectra of dileptons from NJL-box simulations with net-baryon density $\rho_B=0.5\;\mathrm{fm^{-3}}$ and different temperatures $T=100,\,200,\,300$ and $\,400\;\mathrm{MeV}$. The bands are fitted curves within the dilepton invariant mass region $2<M_{ll}<3\,\mathrm{GeV}$.}
	\label{fig:box_dilepton}
\end{figure}

\subsection{Dilepton production from quark matter in a box}
\label{sec:box2}
\begin{table}[htbp]
	\caption{Medium and dilepton temperatures in units of MeV from a quark matter in a box.}
	\centering
	\begin{tabular}{ccccc}
		\hline
		\hline
		$T_\mathrm{Medium}$ & $100$ & $200$ & $300$ & $400$ \\
		\hline
		$T_\mathrm{Dilepton}$ (IMR) & 
		$100$ & $202$ & $307$ & $420$ \\
		\hline
		$T_\mathrm{Dilepton}$ (IMR-I) & 
		$100$ & $205$ & $316$ & $439$ \\
		\hline
		$T_\mathrm{Dilepton}$ (IMR-II)& 
		$103$ & $201$ & $297$ & $401$ \\
		\hline
		\hline
	\end{tabular}
	\label{tab:temp}
\end{table}
We first study dilepton production from an infinite partonic matter at constant temperature and net-baryon density, modeled by imposing periodic boundary conditions on a cubic box of length $L=20$ fm. Figure~\ref{fig:box_dilepton} depicts the invariant mass spectra of dileptons for different medium temperatures of $T_\mathrm{Medium}=100$, 200, 300 and 400 MeV at the fixed net-baryon density $\rho_B=0.5\,\mathrm{fm^{-3}}$. They are obtained from calculations using the NJL transport model described in Sec.~\ref{sec:theory}. To extract the temperature of produced dileptons, the IMR is further divided into two sub-regions of IMR-I ($1<M_{ll}<2\,\mathrm{GeV}$) and IMR-II  ($2<M_{ll}<3\,\mathrm{GeV}$). The values of $T_\mathrm{Dilepton}$ obtained by fitting IMRs of dileptons are shown in Tab.~\ref{tab:temp} with fitting uncertainties less than 1 MeV. The slightly larger $T_\mathrm{Dilepton}$ than $T_\mathrm{Medium}$ is due to difference between the Fermi-Dirac distribution used in the calculations and the Boltzmann distribution used in obtaining Eq.~(\ref{equ:fit}).

\subsection{Dilepton production from an expanding quark matter}
\label{sec:fire2}

For dilepton production from an expanding quark matter, we take the initial net-baryon density and  temperature distributions in the quark matter to have the Woods-Saxon forms,
\begin{align}
	\rho_B(r) &= \frac{\rho_0}
	{1+\exp{\left(\left(r-R_\mathrm{WS}\right)/a_\mathrm{WS}\right)}},\label{equ:rhob}\\
	T(r) &= \frac{T_0}
	{1+\exp{\left(\left(r-R_\mathrm{WS}\right)/a_\mathrm{WS}\right)}},\label{equ:temp}
\end{align}
with a radius $R_\mathrm{WS}=6\;\mathrm{fm}$ and a surface thickness parameter $a_\mathrm{WS}=0.6\;\mathrm{fm}$~\cite{Sun:2020pjz}. The maximum density $\rho_0$ is set to $0.5\,\mathrm{fm^{-3}}$ and the maximum temperature $T_0$ will be varied from 200 MeV to 400 MeV. To account for the effects of collective flow, we take the flow vector $\vec{\beta}$ as a linear function of the position vector $\vec{r}$, i.e.,
\begin{equation}
	\vec{\beta}(\vec{r}) = \frac{\beta_0}{R_{\rm WS}}{\vec r}, \label{equ:beta}
\end{equation}
where $\beta_0$ denotes the maximum value of flow velocity at the initial stage. As for quark matter in a box, we solve the partonic transport model and calculate the dilepton invariant mass spectrum using the method described in Sec.~\ref{sec:theory}.

\begin{figure}[!t]
	\centering
	\includegraphics[width=0.48\textwidth]{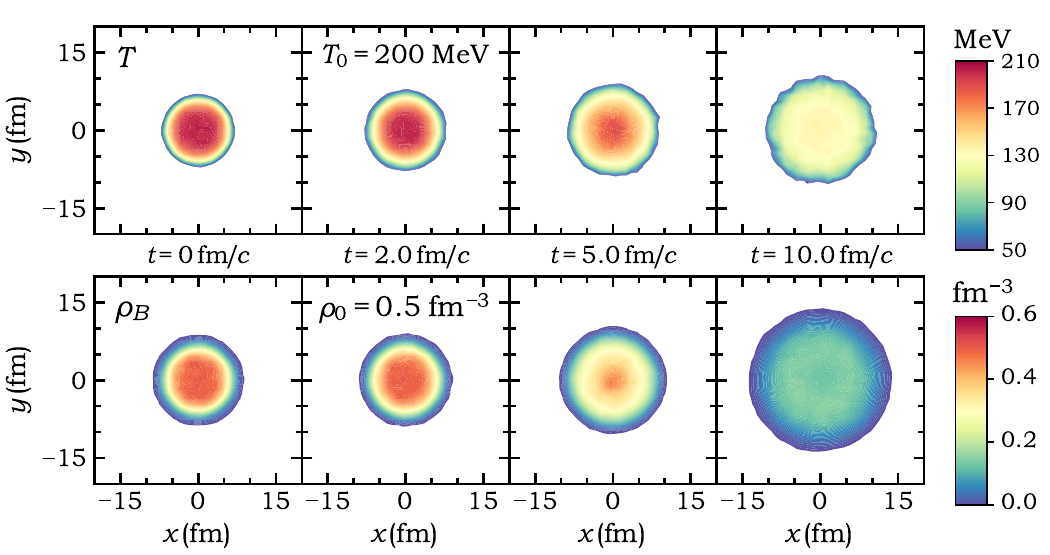}
	\caption{Contour plots of temperature $T$ (first row) and net-baryon density $\rho_B$ (second row) at the $XOY$-plane within $|z|<1$ fm at different times from an expanding quark matter of initial central density net baryon density $\rho_0=0.5\;\mathrm{fm^{-3}}$ and temperature $T_0=200\;\mathrm{MeV}$.}
	\label{fig:temp_rho8}
\end{figure}
Figure~\ref{fig:temp_rho8} depicts the time evolution of the expanding quark matter for an initial temperature $T_0=200\,\mathrm{MeV}$. It is seen that both the central temperature and net-baryon density decrease as the quark matter expands, making the temperature and density distributions in the expanding quark matter more uniform. At about $t=10\,\mathrm{fm\!/}\!c$, the quark matter exhibits, however, a ``bubble'' structure with a lower central density than in the outside region.

\begin{figure}[!t]
	\centering
	\includegraphics[width=0.48\textwidth]{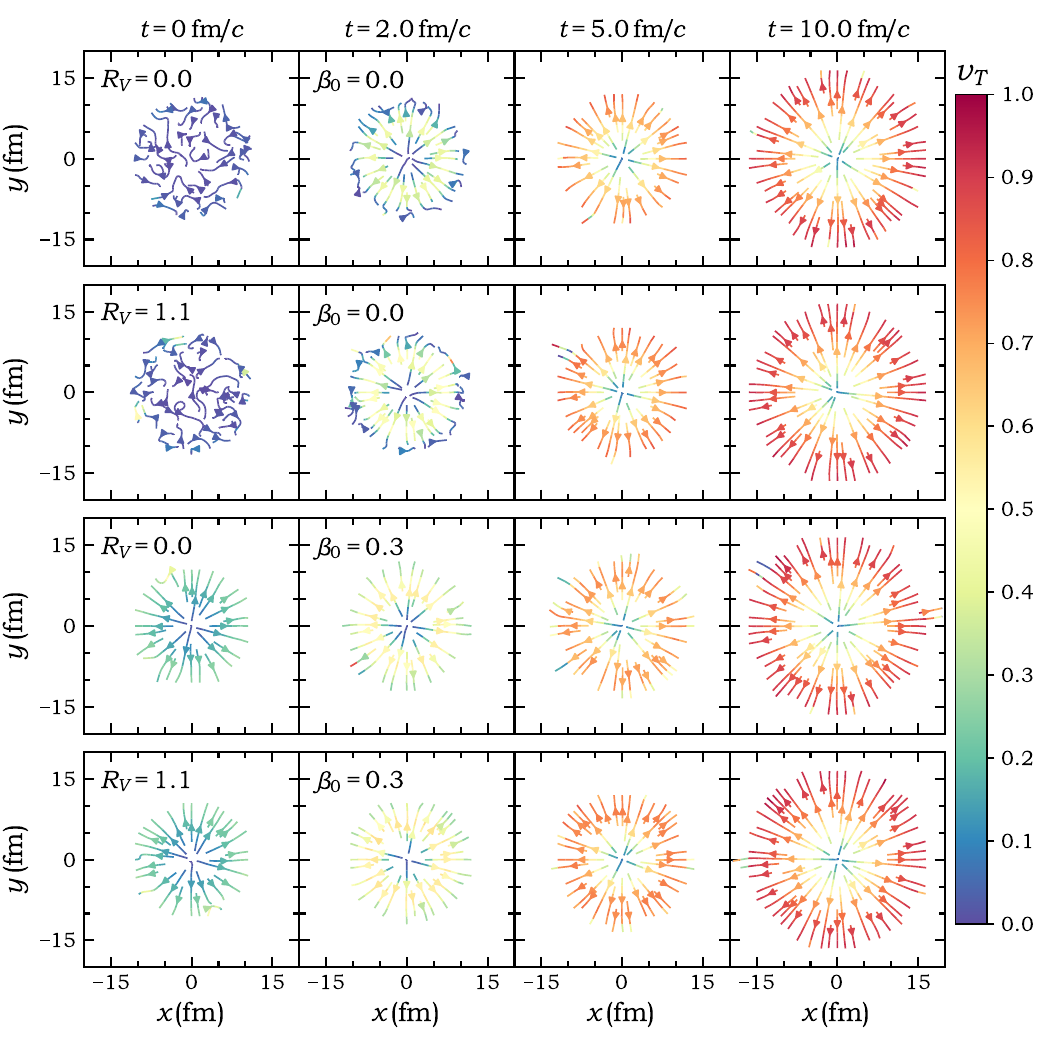}
	\caption{Stream plots of velocities at the $XOY$-plane within $|z|<1$ fm at different times from fireball simulations with same $\rho_0=0.5\;\mathrm{fm^{-3}}$ and $T_0=200\;\mathrm{MeV}$ but different strength of vector coupling $R_V$ and initial flow velocity $\beta_0$.}
	\label{fig:velocity}
\end{figure}

Figure~\ref{fig:velocity} depicts the stream plots of (anti)quark local transverse velocities $v_T=\sqrt{v_x^2+v_y^2}$ in the expanding quark matter for various values of initial flow velocity $\beta_0$ and (anti)quark vector coupling strength $R_V=G_V\!/G_S$. In the absence of initial collective flow, (anti)quark velocity vectors are seen to have random directions. A collective flow is, however, gradually developed by the pressure gradient, resulting in a velocity field at $t=10$ fm/$c$ similar to that of an expanding quark matter of finite initial $\beta_0$, regardless of the values of $R_V$.

\begin{figure}[htbp]
	\centering
	\includegraphics[width=0.45\textwidth]{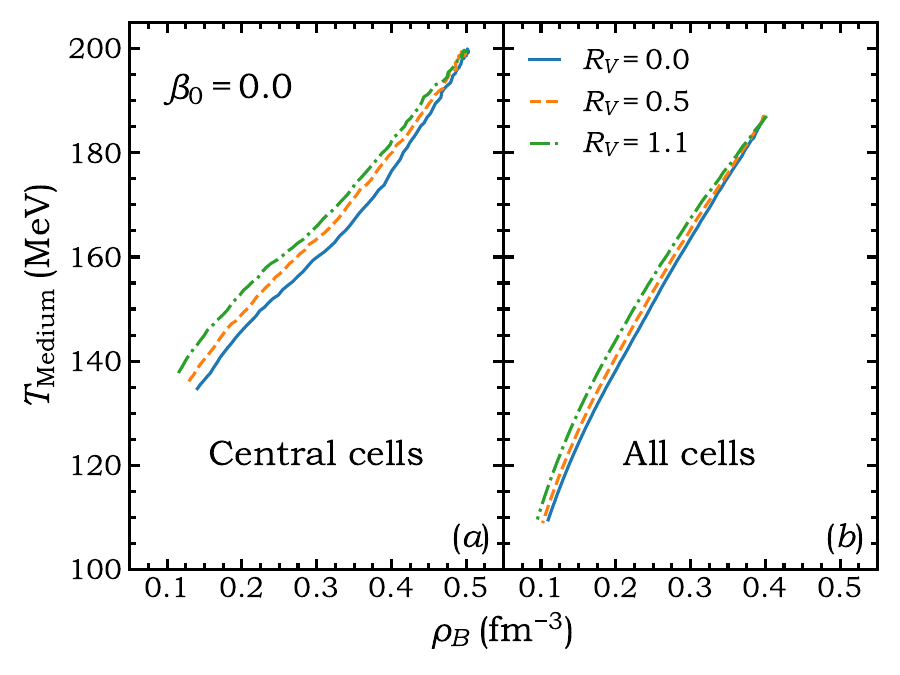}
	\caption{Phase trajectories of an expanding quark matter with initial $T_0=200\;\mathrm{MeV}$, $\rho_0=0.5\;\mathrm{fm^{-3}}$, and $\beta_0=0$ in the temperature and net baryon density plane. Left and right panels correspond to the central region and whole quark matter, respectively.}
	\label{fig:Phase_Rho2_Rv2}
\end{figure}

The dynamics of the expanding quark matter can be described by its phase trajectory in the temperature and net baryon density plane as shown in Fig.~\ref{fig:Phase_Rho2_Rv2} for different values of $R_V$ and zero initial flow $\vec{\beta}=0$. The temperature $T_\mathrm{Medium}$, calculated according to Eq.~\eqref{equ:ther}, and the net-baryon density $\rho_B$ are depicted in the left and right panels, respectively, for the central cells of volume $2\times 2\times 2\,\mathrm{fm^3}$ in the quark matter and the entire quark matter, respectively. These trajectories show that the expansion velocity of the quark matter increases with higher values of $R_V$ as the temperature in the region of same baryon density becomes lower. Although the temperature differences associated with different strengths of $R_V$ increase in the early stages of the quark matter expansion, they remain constant during subsequent evolution. This is mainly due to the effective electric field term in Eq.~\eqref{equ:dkdt}, which dominates over the mass gradient term as the quark density decreases. The differences in the final temperatures obtained between quark matters described with different values of $R_V$ are, however, less than 5 MeV, which is a relatively small value in comparison to the temperature of the quark matter, even at the final stage of the evolution.

\begin{figure}[!t]
	\centering
	\includegraphics[width=0.48\textwidth]{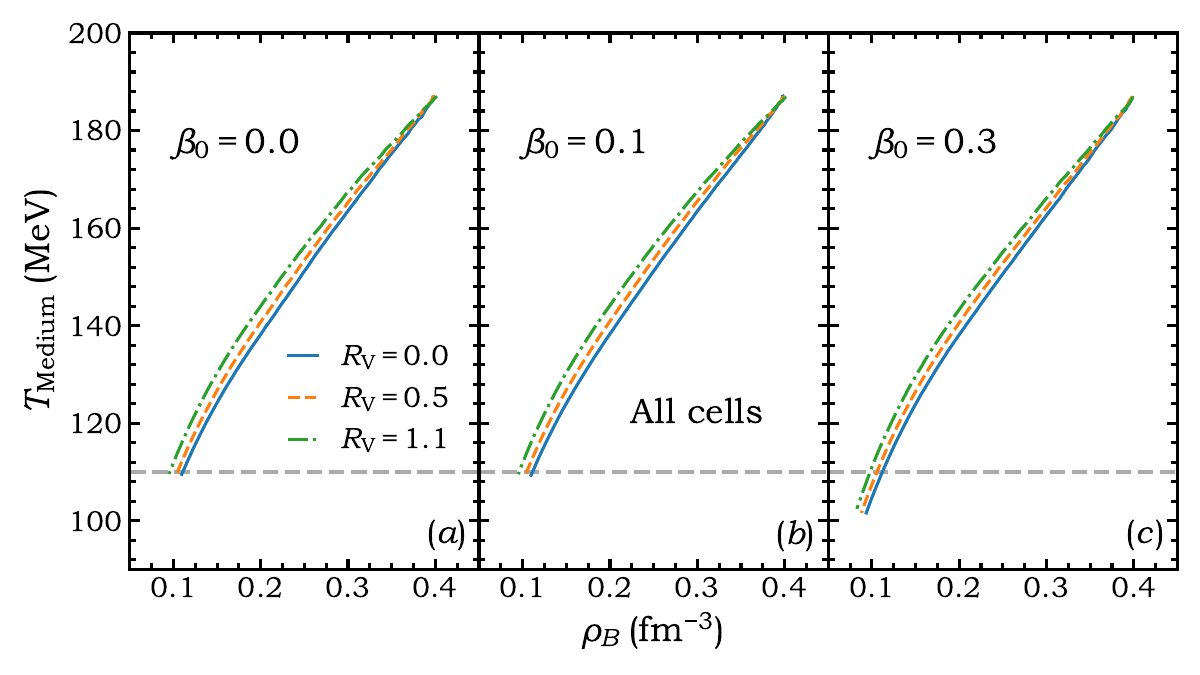}
	\caption{Phase trajectories of quark matter evolution with same $T_0=200\;\mathrm{MeV}$ and $\rho_0=0.5\;\mathrm{fm^{-3}}$ but for different values of vector coupling $R_V$ and initial flow velocity $\beta_0$.}
	\label{fig:Phase_Rho2_flow3}
\end{figure}

Figure~\ref{fig:Phase_Rho2_flow3} illustrates the impact of initial flow on the phase trajectory of the quark matter evolution, with temperatures obtained from the entire quark matter.  Comparisons of the trajectories in this figure reveal that a small $\beta_0 (\leqslant 0.1)$ has negligible effects on the evolution of the quark matter. In contrast, larger $\beta_0$ values can slightly accelerate the decrease of the temperature of the quark matter. Since there is no appreciable dependence of $T_\mathrm{Dilepton}$ on the values of the two parameters $\beta_0$ and $R_V$, they are set to zero in the following studies.

\begin{figure}[htbp]
	\centering
	\includegraphics[width=0.45\textwidth]{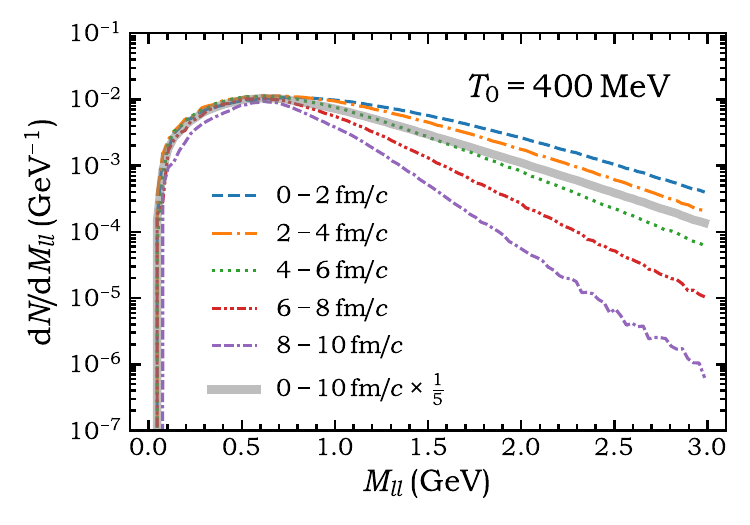}
	\caption{Dilepton invariant mass spectra at different time intervals during the quark matter expansion with an initial temperature $T_0=400$ MeV. The gray band denotes the total dilepton yield scaled by a factor of 5.}
	\label{fig:fireball_time}
\end{figure}

Figure~\ref{fig:fireball_time} depicts dilepton invariant mass spectra during the time intervals of $0-2,\,2-4,\,4-6,\,6-8,\,8-10,\,0-10\,\mathrm{fm}\!/\!c$ of an expanding quark matter with an initial temperature $T_0=400$ MeV. The dilepton yield is seen to decrease as the quark matter expands, with the slope of its invariant mass spectrum dropping rapidly as the temperature decreases. The gray band denotes the total dilepton yield scaled by a factor of 5.   

\begin{figure}[!t]
	\centering
	\includegraphics[width=0.48\textwidth]{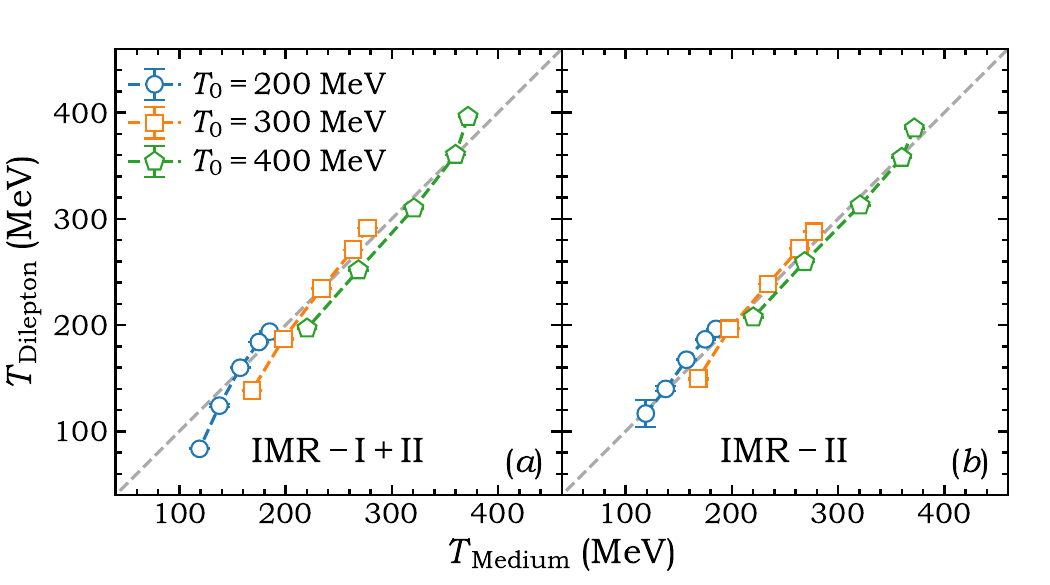}
	\caption{$T_\mathrm{Dilepton}$ versus $T_\mathrm{Medium}$ in an expanding quark matter with an initial baryon density $\rho_0=0.5\;\mathrm{fm^{-3}}$ and different values for the initial temperature $T_0$. The dashed gray line represents equal $T_{\rm Dilepton}$ and $T_{\rm Medium}$. The two panels correspond to $T_\mathrm{Dilepton}$ obtained from fitting different invariant mass region of the dileptons.}
	\label{fig:fire_T4}
\end{figure}

Figure~\ref{fig:fire_T4} shows the temperatures extracted from the dilepton invariant mass spectra in Fig.~\ref{fig:fireball_time} using Eq.~\eqref{equ:fit} versus the medium temperatures $T_\mathrm{Medium}$ calculated using Eq.~\eqref{equ:ther}. It is seen that these two temperatures at different time intervals during the quark matter expansion agree well with each other, especially for temperatures extracted from dilepton spectra at IMR-II as seen in the right panel of Fig.~\ref{fig:fire_T4}. This feature supports the idea of using dilepton production as a thermometer of the quark matter produced in high energy heavy-ion collisions~\cite{Churchill:2023zkk,Churchill:2023vpt}.

\begin{figure}[htbp]
	\centering
	\includegraphics[width=0.41\textwidth]{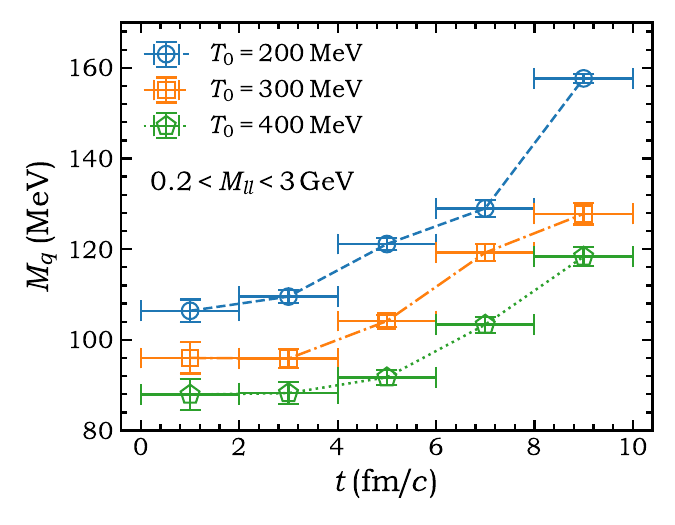}
	\caption{Effective quark mass $M_q$ as a function of the evolution time of an expanding quark matter for different initial temperatures $T_0$.}
	\label{fig:fireball_mass}
\end{figure}

Figure~\ref{fig:fireball_mass} shows the effective quark mass $M_q$ extracted from the dilepton invariant mass spectrum as a function of the quark matter evolution time for various values of the initial temperature $T_0$ and with the fitting range of $M_{ll}$ taken to be $0.2<M_{ll}<3\,\mathrm{GeV}$. It is seen that the effective quark mass gradually increases as the quark matter expands as a result of the spontaneous breaking of chiral symmetry when the baryon density and temperature of the quark matter decrease.

\begin{figure}[!t]
	\centering
	\includegraphics[width=0.45\textwidth]{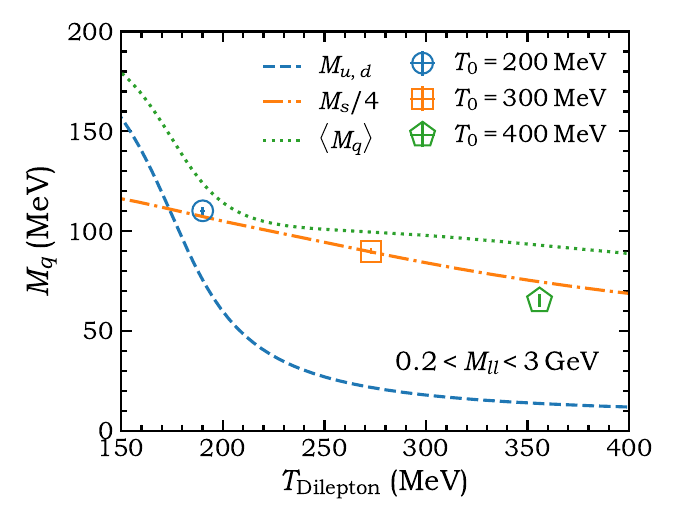}
	\caption{In-medium quark mass as a function of temperature. The three solid symbols denote results extracted from the invariant mass spectra of dileptons produced from an expanding quark matter with three different initial temperatures $T_0$. The lines represent the relation between in-medium quark mass and temperature at a constant baryon number density $\rho_B=0.3\,\mathrm{fm^{-3}}$, obtained by solving the gap equations (Eq.~(\ref{eq:gap})).}
	\label{fig:final}
\end{figure}

In Fig.~\ref{fig:final}, we shown by solid symbols the dependence of effective quark mass $M_q$ on the temperature $T_\text{Dilepton}$ extracted from fitting the dilepton spectra during the quark matter evolution. The results indicate that $M_q$ decreases with increasing $T_\text{dilepton}$, which is consistent with the progressive restoration of the chiral symmetry at higher temperatures. This trend aligns with the predictions obtained by solving the gap equations [Eq.~(\ref{eq:gap})], represented by the colored lines in the figure.  Note that the strange quark mass is divided by a factor of 4, and the average quark mass is denoted by $\langle M_q\rangle$. The results presented in  Figs.~\ref{fig:fireball_mass} and \ref{fig:final}   suggest that dilepton production is a sensitive probe to chiral symmetry restoration in the partonic matter produced in heavy-ion collisions. 

\section{Conclusions}
\label{sec:conclusion}
In summary, we study dilepton production using a partonic transport model with quark mean fields obtained from an extended NJL model. We find that the temperature extracted from the intermediate-mass region of the dilepton spectrum aligns well with the temperature of the partonic matter, which supports the idea that dilepton production can serve as a thermometer for the produced partonic matter. Besides, we find that the extracted quark mass is smaller when the dilepton temperature is higher, suggesting that dilepton production may also serve as a probe to chiral symmetry restoration in high energy heavy-ion collisions. Future studies based on more realistic models to quantify the temperature dependence of chiral symmetry restoration from the invariant mass spectra of dileptons produced in high energy heavy ion collisions, particularly at the RHIC Beam Energy Scan (BES) energies, will be of great interest.

\section*{Acknowledgments}
The authors thank Jie Zhao, Shuai Yang, Zhen Wang, Zao-Chen Ye, Ze-Bo Tang and Wang-Mei Zha for helpful discussions, and Chen Zhong for helping with the computation server. This work was supported in part by the National Natural Science Foundation of China (NSFC) Grant Nos: 12347143, 12375121, 12422509, 12405150, 12147101, 12347106, the Natural Science Basic Research Program of Shaanxi (Program No. 2023-JC-QN-0267 and 2024-JC-YBQN-0043), and the U.S. Department of Energy under Award No. DE-SC0015266.

\bibliographystyle{apsrev4-2}
%

\end{document}